\documentclass[twocolumn,showpacs,preprintnumbers,amsmath,amssymb,
showkeys]{revtex4}

\usepackage{graphicx}
\usepackage{dcolumn}
\usepackage{bm}

\begin{document}

\title{Stars creating a gravitational repulsion}

\author{Igor D. Novikov}
\affiliation{Astro-Space Center of P.N. Lebedev Physical Institute, Profsoyusnaya 84/32, Moscow, Russia 117997.\\
  The Neil's Bohr International Academy, Neil's Bohr Institute, Blegdamsvej 17, DK-2100, Copenhagen, Denmark.\\
  National Research Center Kurchatov Institute, 123182, Moscow, Russia.
}
\author{G.S.Bisnovatyi-Kogan}
\affiliation{Space Research Institute, Profsoyusnaya 84/32, Moscow, Russia 117997.\\
  National Research Nuclear University MEPhI, Kashira Highway, 31, Moscow,
  115409.\\
Moscow Institute of Physics and Technology MIPT,Institutskiy Pereulok, 9, Dolgoprudny, Moscow region, 141701.}

\author{Dmitry I. Novikov}
\affiliation{Astro-Space Center of P.N. Lebedev Physical Institute, Profsoyusnaya 84/32, Moscow, Russia 117997.
}


\begin{abstract}
In the framework of the Theory of General Relativity, models of stars with an unusual equation of state $\rho c^2<0$, $P>0$ where $\rho$ is the mass density and $P$ is the pressure, are constructed. These objects create outside themselves the forces of gravitational repulsion. The equilibrium of such stars is ensured by a non-standard balance of forces. Negative mass density, acting gravitationally on itself, creates an acceleration of the negative mass, directed from the center. Therefore in the absence of pressure such an object tends to expand. At the same time, the positive pressure, which falls just like in ordinary stars from the center to the surface, creates a force directed from the center. This force acts on the negative mass density, which causes acceleration directed the opposite of the acting force, that is to the center of the star. This acceleration balances the gravitational repulsion produced by the negative mass. Thus, in our models gravity and pressure change roles: the negative mass tends to create a gravitational repulsion, while the gradient of the pressure acting on the negative mass tends to compress the star. In this paper, we construct several models of such a star with various equations of state.
\end{abstract}



\maketitle

\section{Introduction}

In the last century, many amazing and unusual objects and phenomena were
discovered, quite unlike those that were known before: white dwarfs,
neutron stars, black holes, dark matter and dark energy.
These discoveries were preceded by exotic theoretical predictions.
In many cases the scientific community were very skeptical about such
predictions. A striking example of this is the discovery of black holes
and the accelerated expansion of the Universe due to the dark energy. Sometimes predictions of this kind are not justified for a long time, but there are hopes of finding such objects in the future. An example of this kind is the hypothesis of wormholes [1,2,3].

In this article, we consider models of objects consisting of a substance
with an unusual equation of state. In particular, the negative energy density
is under consideration. The appeal to this possibility is connected, of course, with the discovery of gravitational repulsion forces that make the universe
expand faster as it was during the inflation and in the modern era. The source of gravitational repulsion in cosmology is the negative pressure $P$.
However, in our case, anti-gravity is due to negative energy density,
rather than negative pressure:

\begin{equation}
  P>0,\hspace{1cm} \rho c^2< 0.
\end{equation}
The possibility of such a condition has been considered in theory long
time ago. This consideration is relevant to the cosmological constant
$\Lambda$, which can be positive or negative and is interpreted as the components of the stress-energy tensor of the vacuum [4,5,6,7,8]. The negative value of
$\Lambda$ corresponds to inequality (1).

Another example of an exotic matter with the equation of state,
that corresponds to this inequality is the scalar field with the
negative energy density [9,10]. Such an equation of state is widely used
in the theory of wormholes [11].

Models of stars with an unusual equation of state were repeatedly considered earlier [12,13,14]. As indicated above, we refer here to the equation of state satisfying (1) specifically in connection with the consideration of the problem of anti-gravity in the Universe [15]. We emphasize that a ball of finite radius filled with a matter with the equation of state corresponding to dark energy in today's Universe ($P\approx -\rho c^2<0$) creates ordinary gravity outside itself in the vacuum, not anti-gravitation. The question arises: can an object exist that create anti-gravity in the space beyond it's border? A positive answer to this question has long been known: such objects are the entrances to wormholes in many models [16,17].
Our article is devoted to the study of the question of whether there can be an object with the usual spherical topology, which creates a gravitational repulsion outside itself.

We will consider models with $\rho<0$ and $P>0$. The motive for considering such models is a very simple example of the mechanical interaction of two bodies with positive and negative masses. In this example, the ball of mass $m_1>0$ moves with the speed $\vec{v}_1$ in the direction of the resting ball with the mass $m_2<0$ [18]. If $|m_2|>m_1$, then after the collision both balls will move in the direction opposite to $\vec{v}_1$. If we denote the velocities of two balls after the collision as $v_1',v_2'$, then we always have $\mid v_1'\mid >\mid v_2'\mid$. A similar example was also considered in [19] for the relativistic linear motion of two particles with masses of opposite signs and a small difference between their absolute values.

In case of a star with $\rho<0$ and $P>0$, the negative matter density creates a gravitational force directed toward the center. This force, acting on a negative mass, creates an acceleration directed from the center of the star. Positive pressure falling down from the center towards the surface creates a force directed from the center. This force, acting on the negative mass, creates an acceleration directed against the acting force, (as in the example with the balls) that is to the center. Thus, gravity and pressure act in opposite directions, balancing each other. Note, that they act in the directions opposite to those in which they act in an ordinary star with a positive matter density.

The paper is organized as follows. Section II gives the equation of equilibrium of a star, which we transform using dimensionless quantities. In Section III we consider the models of a star with a given equation of state and models with a given density profile. Finally in Section IV we make our conclusive remarks.

\section{The equilibrium of the star}
The equation of equilibrium for the spherical star in General Relativity can be
written in the following form [7]:

\begin{equation}
    \frac{dP}{dr}=-G\frac{(\rho c^2+P)(M_rc^2+4\pi Pr^3)}
         {r^2c^4-2GM_rrc^2},
\end{equation}
where $G$ is the gravitational constant, $c$ is the speed of light,
$r$ is the radial coordinate ($r^2=A/4\pi$, where $A$ is the total area of the 2-sphere), $P$ is the pressure,
$\rho$ is the density and $M_r$ is:
\begin{equation}
    M_r=4\pi\int\limits_0^r\rho(\tau)\tau^2d\tau,
\end{equation}
For further consideration, we denote the absolute value of the density at
the center of the star as $\rho_c$ and use the dimensionless quantities
$\theta$, $w$, $x$:
\begin{equation}
  \rho=\rho_c\theta,\hspace{0.2cm}P=\rho_cc^2w,
  \hspace{0.2cm}r=Rx,
  \hspace{0.2cm}R^2=\frac{c^2}{4\pi G\rho_c},
\end{equation}
In our models we always have $\theta<0$ and $w>0$, so that the equation of
state satisfies (1).

The equation (2) in dimensionless units (4) looks as follows:

\begin{equation}
    \frac{dw}{dx}=-\frac{(\theta+w)(I_x+x^3w)}
         {x^2-2xI_x},
\end{equation}

and

\begin{equation}
I_x=\int\limits_0^x\theta(\tau)\tau^2d\tau<0
\end{equation}
is the dimensionless mass, which  corresponds to the expression (3).

\section{Models of a star creating anti-gravitation}

\begin{figure*}[tbh]
  \includegraphics[width=0.9\textwidth]{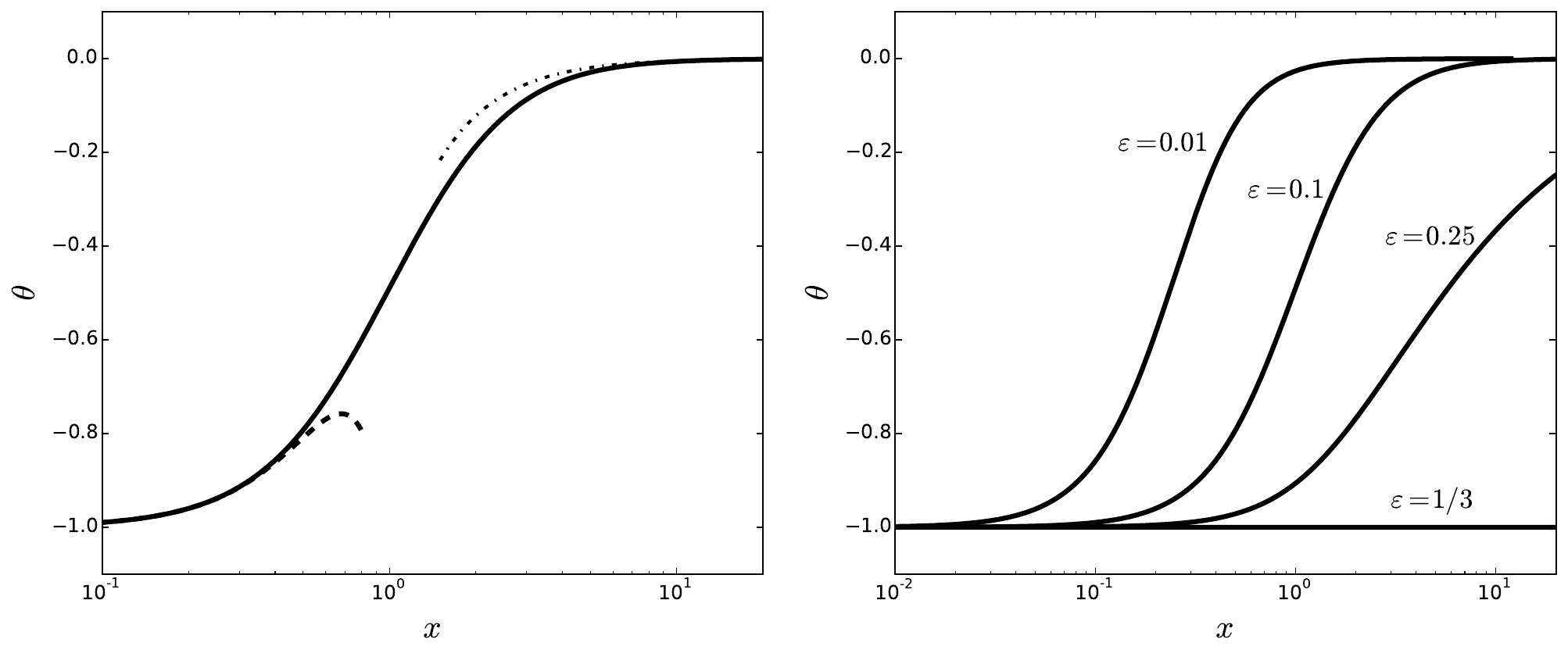}
  \caption{The density profile as a function of the coordinate $x$
    for the model with the equation of state $w=-\varepsilon\theta$.
    Left panel: the density profile for $\varepsilon=0.1$ (solid line),
    together with two asymptotics: the dashed line for
    $x\ll\frac{1}{\sqrt{d_2}}$ and the dotted line for
      $x\gg\frac{1}{\sqrt{d_2}}$. Right panel:
    the density profile for
    different values of $\varepsilon$.}
\end{figure*}
\begin{figure*}[tbh]
  \includegraphics[width=0.9\textwidth]{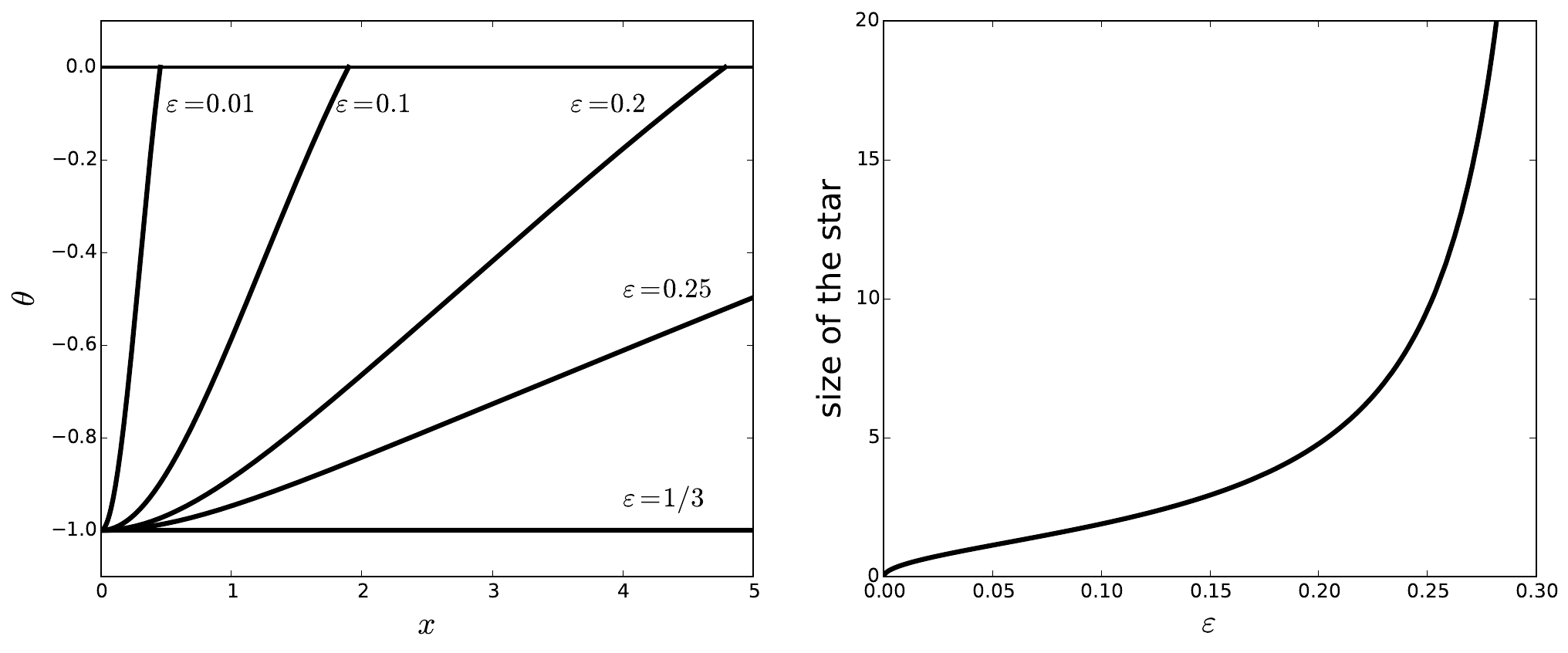}
  \caption{The model with the equation of state $w=\varepsilon\theta^2$.
    Left panel:
    density as a function of the coordinate $x$ for different values
    of $\varepsilon$. Right panel: size of the star as a function
    of $\varepsilon$.}
\end{figure*}

In this Section we consider two types of models of a star with negative mass density: models with a given equation of state and models with a given density profile. In our paper, when specifying the equation of state, we confine ourselves to two models: the linear and quadratic dependence of pressure on density. As for models with a given density profile, we consider here the model of a star filled with matter of constant density and a model with a parabolic dependence of the density on the coordinate.

\subsection{The models with the given equation of state}

\begin{center}
{\it The model with the equation of state
  \boldmath${w=-\varepsilon\theta},\hspace{0.5cm}\varepsilon>0$\unboldmath.}
\end{center}

The motive for applying this equation of state is the fact that the linear relation between pressure and density has been considered in classical works devoted to the study of the star equilibrium for a positive mass density.

The equation (5) for the object with the negative matter density and under condition
$w=-\varepsilon\theta$, where $\varepsilon$ is the positive constant, takes
the following form:

\begin{equation}
  \theta'=\theta\cdot\frac{1-\varepsilon}{\varepsilon}\cdot
  \frac{I_x-\varepsilon x^3\theta}
       {x^2-2xI_x},\hspace{0.7cm}\theta'=\frac{d\theta}{dx}
\end{equation}

We should mention first, that this equation has a simple analytical
solution (a similar analytical solution was found in [20, 21] for a positive mass density and positive pressure):

\begin{equation}
  \theta=-\frac{2\varepsilon}{(\varepsilon-3)^2-8}\cdot\frac{1}{x^2},
  \hspace{0.3cm}for\hspace{0.3cm}\varepsilon<3-2\sqrt{2}\approx 0.17.
\end{equation}

Such a solution corresponds to infinite density in the
center of the star. In order to satisfy the finite density in the center
we should integrate the equation (7)
with the boundary condition $\theta(0)=-1$,\hspace{0.5cm}$\theta'(0)=0$.
The solution in the vicinity of the center can be found analytically
by representing $\theta$ in the form of Taylor series with even terms only:

\begin{equation}
  \theta=-1+d_2x^2+d_4x^4+...,\hspace{0.7cm}
  d_n=\left.\frac{1}{n!}\frac{d^n\theta}{dx^n}\right|_0.
\end{equation}
We use here even terms to make sure, that there is no any weird feature in
the density or pressure profile around the center of the star. Thus, $\theta$ and $w$ are
continuous functions, all their derivatives have no discontinuities and
our solution is spherically symmetrical.
Substituting expression (9) in equation (7) we have expressions for $d_n$:
\begin{equation}
  \begin{array}{l}
    \vspace{0.3cm}
  d_2=\frac{(1-\varepsilon)}{2\epsilon}\left(\frac{1}{3}-\varepsilon\right),\\
  d_4=\left[\frac{1-\varepsilon}{2\epsilon}
    \left(\varepsilon-\frac{4}{15}\right)-\frac{1}{3}\right]d_2.
  \end{array}
\end{equation}

In order to make sure, that density is changing monotonically from
$-1$ to $0$ in the direction from the center, the equation of state should
satisfy the condition $\varepsilon<1/3$. The
asymptotic solution for large $x$ can either be the expression (8)
if $\varepsilon<3-2\sqrt{2}$ or if $\varepsilon>3-2\sqrt{2}$ then
$\theta\sim x^\gamma$, $\gamma>-2$.
Finally we have the following asymptotics:
\begin{equation}
  \begin{array}{l}
     \vspace{0.3cm}
     \theta \approx -1+d_2x^2+d_4x^4,\hspace{0.2cm}x\ll\frac{1}{\sqrt{d_2}},
     \hspace{0.2cm}
  \varepsilon<\frac{1}{3};\\
  \vspace{0.3cm}
  \theta\approx -\frac{2\varepsilon}{(\varepsilon-3)^2-8}\cdot\frac{1}{x^2},
  \hspace{0.2cm}x\gg\frac{1}{\sqrt{d_2}},\hspace{0.2cm}
  \varepsilon<3-2\sqrt{2};\\
  \theta\approx Cx^\gamma,\hspace{0.2cm}\gamma=
  \frac{(3\varepsilon-1)(1-\varepsilon)}{\varepsilon(1+\varepsilon)},
  \hspace{0.2cm}x\gg\frac{1}{\sqrt{d_2}},\\
  3-2\sqrt{2}<\varepsilon<\frac{1}{3},
  \end{array}
\end{equation}
where $C<0$ is some negative constant.
Results of our numerical integration of equation (7) for different
values of $\varepsilon$ is shown in Fig 1.

It should be mentioned that for the equation of state under consideration, when the pressure is proportional to the density, the mass of the star turns out to be infinite, since the integral
$I_{\infty}=\int\limits_0^{\infty}\tau^2\theta d\tau$ does not converge. In order to avoid mass infinity, below we consider a model of a star with a
different equation of state, which gives us the finite size of the star and hence its finite mass.

\begin{center}
{\it The model with the equation of state
  \boldmath${w=\varepsilon\theta^2},\hspace{0.5cm}\varepsilon>0,
  \hspace{0.5cm}\theta<0$\unboldmath.}
\end{center}
In case of the equation of state $w=\varepsilon\theta^2$ the expression
(5) becomes:
\begin{equation}
  \frac{d\theta}{dx}=-\frac{1+\varepsilon\theta}{2\varepsilon}\cdot
  \frac{I_x+\varepsilon x^3\theta^2}{x^2-2xI_x}.
\end{equation}
As can easily be seen, if $\theta=0$ then
the density derivative with respect to the coordinate when $\theta=0$ is a finite positive constant. This means, that the model of the star with
finite size and mass can be constructed:

\begin{equation}
  \left.\frac{d\theta}{dx}\right|_{x=a}=-\frac{1}{2\varepsilon}\frac{I_a}
       {a^2-2aI_a},\hspace{0.2cm}\theta(a)=0,\hspace{0.2cm}I_a<0.
\end{equation}
Here and below by $a$ we denote the size of the star. We define
$a$ as a coordinate at which the pressure becomes zero: $w(a)=0$. In this
particular case since $w=\varepsilon\theta^2$, both pressure and
its derivative at $a$ are zero: $w(a)=\frac{dw}{dx}\mid_{x=a}=0$.

Analogously to the previous case one can find the asymptotic solution of (12)
for small $x$:
\begin{equation}
  \theta\approx -1+\frac{(1-\varepsilon)}{4\varepsilon}
  \left(\frac{1}{3}-\varepsilon\right)\cdot x^2.
\end{equation}

The numerical solution of (12) is shown in Fig 2. $\theta(x)$ is growing
monotonically and unlike the previous case eventually becomes zero at $x=a$, $\theta(a)=0$, where $a$ is the size of the star. In Fig 2 we also show the dependence
of the size of the star on $\varepsilon$. As we can see this size growing
as $\varepsilon$ increases and becomes infinite for $\varepsilon=1/3$ because
in this case the equation (12) has the constant solution $\theta=-1$.

\subsection{The models with a given density profile}

Motivation to consider models with a given density profile is the fact that the equation of state can vary depending on the distance from the center of the star.
In this subsection we consider the models with a given monotonic function
$\theta$, $-1<\theta<0$ at the range $0<x<\infty$. In this case the pressure
$w$ satisfies the equilibrium equation (5) and eventually becomes zero
at $x=a$, where $a$ is the size of the star.

\begin{center}
{\it The constant density model.}
\end{center}

We start our analysis with the model of the star with constant density as a special case of the models with a given density profile. In such a model
the density $\theta=-1$ and the equation of equilibrium (5)
takes a particularly simple form:
\begin{equation}
 \frac{dw}{dx}=-x\cdot\frac{(w-1)(w-\frac{1}{3})}{1+\frac{2}{3}x^2}
\end{equation}
This equation can be easily integrated and has the following analytical
solution:
{\Large
\begin{equation}
  \begin{array}{l}
    \vspace{0.5cm}
    w=\frac{(3w_c-1)\sqrt{1+\frac{2}{3}x^2}-(w_c-1)}
    {(3w_c-1)\sqrt{1+\frac{2}{3}x^2}-3(w_c-1)},\\
    a^2=-\frac{3w_c(4w_c-2)}{(3w_c-1)^2},\hspace{1cm}\varepsilon<\frac{1}{3}
  \end{array}
\end{equation}
\par}
where $w_c$ is the central pressure and it is easy to see, that $w$
becomes zero at $x=a$.
In this model the ratio $w/\theta$ is changing monotonically:
\begin{equation}
  -\frac{1}{3}<\frac{w}{\theta}=-w<0,\hspace{0.2cm}as\hspace{0.2cm}
  0<x<a,\hspace{0.2cm}0<a<\infty.
\end{equation}

The equilibrium equation (5) for the constant density in case of positive mass density was integrated and the solution was analyzed in [22, 23]. For this case, there is a restriction, which looks as :

\begin{equation}
I_a<\frac{4}{9}a.
\end{equation}
It's important to note, that there is no such a restriction for the negative
mass density.

\begin{figure}[tbh]
  \includegraphics[width=1\columnwidth]{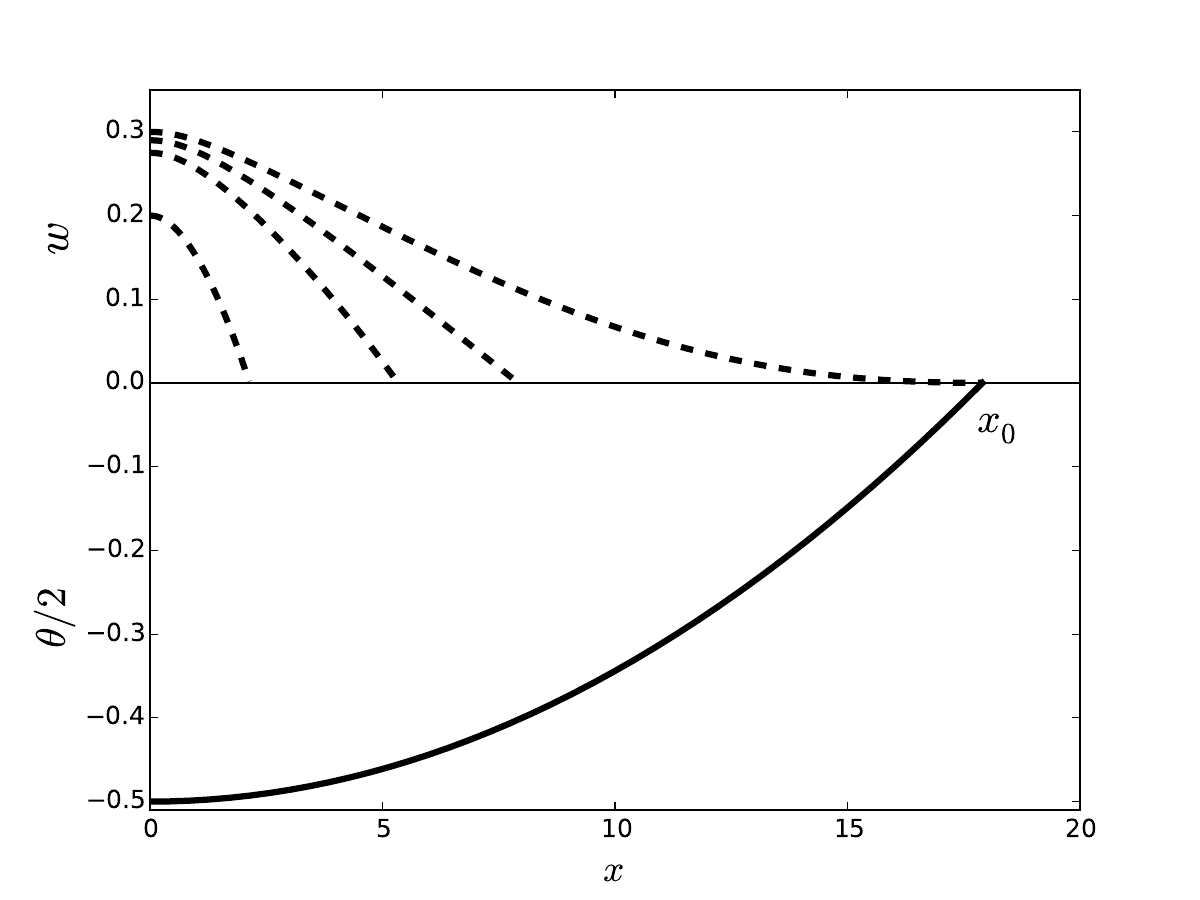}
  \caption{Pressure as a function of the coordinate for a given density profile. Solid line: density. Shaded lines: pressure for different values of the central pressure $w_c=w(0)$. For $w_c=0.3$,
    the parameter $x_0=17.9$ gives the example in which the pressure and
    density simultaneously turn to zero.}
\end{figure}

\begin{figure*}[bth]
  \includegraphics[width=1\textwidth]{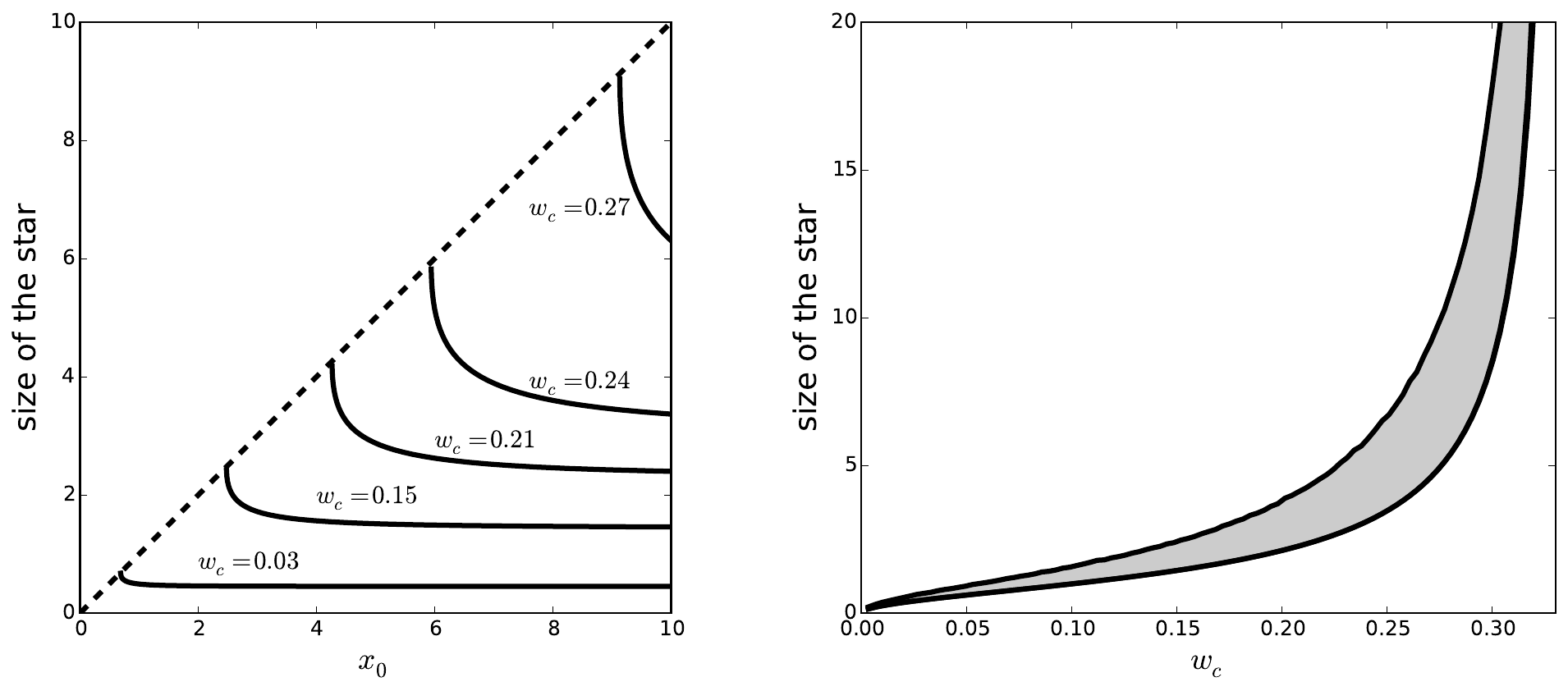}
  \caption{The model with the given density profile. Left panel: dependence of the size of the star on the parameter $x_0$ for
    different values of the central pressure. The dashed line shows the values of
    $x_0$ at which the density and pressure simultaneously turn to zero ($w(x_0)=\theta(x_0)=0$). For $x_0\rightarrow\infty$, the size of the star asymptotically tends to a size determined by the model with constant density. Right panel: the size of a star as a function of the central pressure. The lower curve determines the size of the star for the constant density model. The upper curve corresponds to a solution in which the density and pressure become zero at the same coordinate $x_0$. The shaded zone corresponds to the range of admissible values of the star size.
}
\end{figure*}

\begin{center}
{\it Models with the parabolic density profile}
\end{center}

Here we consider a more general but still extremely simple case of the density
profile, namely the parabolic shape for $\theta$:

\begin{equation}
\theta=-1+\left(\frac{x}{x_0}\right)^2
\end{equation}
Note, that in case of $x_0\rightarrow\infty$ this model reduces to a constant density model (see above).

In order to find the solution we should substitute (19) to (5)
and define the boundary condition at $x=0$ as $w(0)=w_c$ and
$\frac{dw}{dx}\mid_{x=0}=0$. After
the substitution we get the differential equation:

\begin{equation}
  \frac{dw}{dx}=-x\cdot\frac{\left[w-1+\left(\frac{x}{x_0}\right)^2\right]
    \left[w-\frac{1}{3}+\frac{1}{5}\left(\frac{x}{x_0}\right)^2\right]}
       {1+\frac{2}{3}x^2-\frac{2}{5}x_0^2\left(\frac{x}{x_0}\right)^4}.
\end{equation}

Analogously to Subsection III A, we represent the pressure in the vicinity of the star center in the form of the Taylor series with even terms:

\begin{equation}
 w=w_c+p_2x^2+p_4x^4+...,\hspace{0.5cm}
  p_n=\left.\frac{1}{n!}\frac{d^nw}{dx^n}\right|_0
\end{equation}
and for coefficients $p_n$ we have:
\begin{equation}
  \begin{array}{l}
  \vspace{0.3cm}
  p_2=-\frac{1}{2}(w_c-1)\left(w_c-\frac{1}{3}\right),\\
  \vspace{0.3cm}
  p_4=-\frac{1}{4}\left[\left(w_c-1\right)\left(p_2+\frac{1}{5x_0^2}\right)
    \right.+\\
    +\left.\left(w_c-\frac{1}{3}\right)
    \left(p_2+\frac{1}{x_0^2}\right)+\frac{4}{3}p_2\right].
  \end{array}
\end{equation}
Therefore, expressions (21,22) give us the analytical solution of eq (20) for
$x\ll\frac{1}{\sqrt{p_2}}$. Results of our numerical integration of this
equation are given in Fig 3. If $x_0\gg\frac{1}{\sqrt{p_2}}$ then
the solution becomes the same as for the model with constant density profile,
see eq. (15). In this case the size of the star (the coordinate
$x=a$, at which the pressure becomes zero) is determined by the central
pressure only and this size is much less then $x_0$. The size of the star increases with increasing central pressure and for any given $x_0$ there exists a
maximum central pressure $w_c$ at which the density and pressure go to zero for the same $x=x_0$: $w(x_0)=\theta(x_0)=0$.

In Fig 4 we demonstrate the dependence of the star size $a$ on $x_0$ and on the central pressure $w_c$. For a fixed value of $w_c$, there are two asymptotics for the star size as a function of $x_0$. For large $x_0$, the size of the star is determined by eq (15). As $x_0$ decreases, the size of the star increases and eventually reaches the maximum possible value when $w(a)=\theta(a)=0$, $a=x_0$.

\section{Conclusive remarks}

As it was stated in the introduction, anti-gravity in cosmology, which causes
the accelerated expansion of the Universe, was widely discussed during the last
quarter of the century. However the source of
anti-gravitation in the cosmological models  was negative pressure [15].
The energy density was assumed to be positive, and such an energy was named in
cosmology as the dark energy.
As it was stressed in [15,17], an isolated body consisting of dark energy
creates outside an
attraction and not anti-gravity. At the same time, a model of the body was
indicated in [17], which creates outside itself the gravitational repulsion.
Such a body is one of the entrances to a wormhole with a massless scalar field
with negative energy density analyzed in [17].
In this paper, we constructed several models of isolated objects of the star
type, creating an anti-gravity. Any test bodies with positive or negative mass outside
such an object will be accelerated away from it.

We did not consider here the issues of the stability of the solutions obtained and did not touch upon the problem whether such objects can have anything to do with the real Universe. Other physical limitations were not considered as well. We would like to mention only, that the standard conditions for the homological stability of a star with a power equation of state $P=\varepsilon\rho^\gamma$ and negative mass density $\rho<0$ will be inverse to those that exist for an ordinary star. Namely, it looks as $\gamma<4/3$. It's nesessary to mention also
that in case $\rho<0$ there is no gravitational radius in the spherical solution.

We also emphasize that the general problem of the positivity of the energy and the positive nature of its radiation is analyzed in [23] pp. 285-295.

\begin{acknowledgments}
We wish to acknowledge the support by the programm of the Presidium of the Russian Academy of Sciences $\Pi$ -28. The work of GSBK was partially supported by RFBR grants 17-02-00760, 18-02-00619 and RAN Program N28 "Astrophysical objects as cosmic laboratories".
\end{acknowledgments}


\end{document}